\documentclass[12pt,letterpaper]{article}
\usepackage{amssymb}
\usepackage{amsmath}
\usepackage{amscd}
\usepackage{graphicx}
\usepackage{tikz}

\def\be{\begin{equation}}
\def\ee{\end{equation}}
\def\bea{\begin{eqnarray}}
\def\eea{\end{eqnarray}}
\def\bse{\begin{subequations}}
\def\ese{\end{subequations}}

\def\U{\Upsilon}

\def\vCent#1{\vcenter{\hbox{\hss#1\hss}}}

\def\pa{\partial}      
\newcommand{\bm}[1]{\mbox{\boldmath$#1$}}

\def\a{{\alpha}}

\def\g{{\gamma}}
\def\s{{\sigma}}
\def\ad{{\dot{\alpha}}}

\def\gd{{\dot{\gamma}}}

\def\D{{\rm D}}
\def\Dd{{\bar{\rm D}}}

\def\[{\left[}
\def\]{\right]}

\font\ro=cmsy10                          
\def\kcr{{\hbox{\ro \char'170}}}                
\def\ktl{{\hbox{\ro \char'170}}}        
\def\ktr{{\hbox{\ro \char'170}}}        
\def\kbl{{\hbox{\ro \char'170}}}        
\def\kbr{{\hbox{\ro \char'170}}}        

\parskip=\medskipamount                 
\thicklines                         

\newskip\humongous \humongous=0pt plus 1000pt minus 1000pt
\def\caja{\mathsurround=0pt}
\def\eqalign#1{\,\vcenter{\openup2\jot \caja
        \ialign{\strut \hfil$\displaystyle{##}$&$
        \displaystyle{{}##}$\hfil\crcr#1\crcr}}\,}
\newif\ifdtup

\def\border{                                            
        \setlength{\unitlength}{1mm}
        \newcount\xco
        \newcount\yco
        \xco=-21
        \yco=12
        \begin{picture}(140,0)
        \put(\xco,\yco){$\ktl$}
        \advance\yco by-1
        {\loop
        \put(\xco,\yco){$\kcr$}
        \advance\yco by-2
        \ifnum\yco>-240
        \repeat
        \put(\xco,\yco){$\kbl$}}
        \xco=158
        \yco=12
        \put(\xco,\yco){$\ktr$}
        \advance\yco by-1
        {\loop
        \put(\xco,\yco){$\kcr$}
        \advance\yco by-2
        \ifnum\yco>-240
        \repeat
        \put(\xco,\yco){$\kbr$}}
        \put(-20,13){\tiny **University of Maryland * Center for String and
         Particle  Theory* Physics Department***University of Maryland *Center
        for String and Particle  Theory** }
        \put(-20,-241.5){\tiny **University of Maryland * Center for String and
         Particle  Theory* Physics Department***University of Maryland *Center
        for String and Particle  Theory** }
        \end{picture}
        \par\vskip-8mm}

\def\headpic{                                           
        \indent
        \setlength{\unitlength}{.4mm}
        \thinlines
        \par
        \begin{picture}(29,16)
        \put(165,16){\line(1,0){4}}
        \put(170,16){\line(1,0){4}}
        \put(180,16){\line(1,0){4}}
        \put(175,0){\line(1,0){4}}
        \put(180,0){\line(1,0){4}}
        \put(185,0){\line(1,0){4}}
        \put(169,0){\line(0,1){16}}
        \put(170,0){\line(0,1){16}}
        \put(179,0){\line(0,1){16}}
        \put(180,0){\line(0,1){16}}
        \put(184,0){\line(0,1){16}}
        \put(185,0){\line(0,1){16}}
        \put(169,16){\oval(8,32)[bl]}
        \put(170,16){\oval(8,32)[br]}
        \put(179,0){\oval(8,32)[tl]}
        \put(185,0){\oval(8,32)[tr]}
        \end{picture}
        \par\vskip-6.5mm
        \thicklines}
\def\endtitle{\end{quotation}\newpage}                  

\begin{document}

\border\headpic {\hbox to\hsize{March 2011 \hfill
{UMDEPP 11-004}}}
\par
{$~$ \hfill
{MIT-CTP-4221}}
\par
{$~$ \hfill
{hep-th/1103.3564}
}
\par

\setlength{\oddsidemargin}{0.3in}
\setlength{\evensidemargin}{-0.3in}
\begin{center}
\vglue .10in
{\large\bf A Codicil To Massless Gauge Superfields of \\Higher Half-Odd Integer Superspins \footnote
{Supported in part  by National Science Foundation Grant
PHY-0354401.}\  }
\\[.5in]

S.\, James Gates, Jr.\footnote{gatess@wam.umd.edu}
and Konstantinos Koutrolikos\footnote{koutrol@umd.edu}
\\[0.2in]

{\it Center for String and Particle Theory\\
Department of Physics, University of Maryland\\
College Park, MD 20742-4111 USA}\\[2.8in]

{\bf ABSTRACT}\\[.01in]
\end{center}
\begin{quotation}
{We study theories of 4D, $\cal N$ $=$ 1 supersymmetric massless, arbitrary higher half odd-integer 
superspins.  A new series of such theories is found to exist for arbitrary superspin $Y$  ($Y$ $=$ $s$ $+$
1/2 for any integer $s$). The lowest member ($s$ $=$ 1) of the series is the original off-shell 
formulation of 4D, $\cal N$ $=$ 1 supergravity first presented by Breitenlohner in 1977.
}
${~~~}$ \newline ${~~~}$ \newline
PACS: 04.65.+e
\endtitle


\section{Introduction}

$~~~$ The state-of-the-art understanding on the subject of higher spin supersymmetric 
multiplets was established in a work by Kuzenko, Postnikov, and Sibiryakov \cite{Kuzenko:1993jp}. 
In fact, they established two such formulations for each and every possible value of the superspin
$Y$.
These formulations are based on the introduction of constrained compensating superfields. The
goal of this work is to re-examine these schemes in order to be able to reproduce their results and, 
if possible, to discover new formulations in the case of half odd superspins. This is exactly what
will happen in the following. Their results will emerge naturally from our algorithm as a possible 
way a theory of higher, half odd massless superspins can be formulated. 

In an accompany paper \cite{Gates:2011qb} devoted to the study of massless 4D, $\cal N$ $=$ 1 higher integer 
superspins, we developed an algorithm that was able to do two things:
\begin{itemize}
\item generate all known results for massless 4D, $\cal N$ $=$ 1 higher integer superspins 
up to that point, and
\item introduce a new formulation of the theory.
\end{itemize}
After the success of this algorithm in the investigation of higher, integer superspins we would 
like to apply a similar way of thinking in the case of half odd superspins.

The conceptual backbone of the method followed, can be summarized as following:\\
$~~~~~$ Step 1) find the main physical superfield\footnote{This fixes the
index structure, mass dimensions and says something about reality of this\\
$~~~~~~$  main physical superfield. Within the context of
supergravity, this main physical superfield \\
$~~~~~~$
 is known as the `superconformal submultiplet.'}, 
that will be used to \newline 
$~~~~~~~~~\,~~~~~~$ construct the theory,\\
$~~~~~$ Step 2) find the most general free action which is quadratic to this \newline
$~~~~~~~~~\,~~~~~~$
superfield,\\
$~~~~~$ Step 3) find the gauge transformation of the main superfield,\\
$~~~~~$ Step 4) find the type of superfield(s) we have to introduce as  \newline
$~~~~~~~~~\,~~~~~~$
compensators,\\
$~~~~~$ Step 5) find the possible gauge transformation{s} of the com-
 \newline
$~~~~~~~~~\,~~~~~~$
pensators which on-shell give just the degrees of \newline
$~~~~~~~~~\,~~~~~~$
freedom needed, and \\
$~~~~~$ Step 6) check invariances of the action with respect to all trans- 
 \newline
$~~~~~~~~~\,~~~~~~$ formations.

\section{General Action and Gauge Transformations}

$~~~$ The goal is to develop a theory for massless half odd superspin $Y$ $=$ $s+\frac{1}
{2}$, for integers $s$. This means the highest superspin projection operator acting on the `main 
superfield'  used to develop the theory must generate an object with an odd number of indices 
($2s+1$).   As suggested by supergravity theory, the fundamental superfield for this theory should 
be a bosonic superfield with an even number of indices, $s$ undotted and $s$ dotted ($H_{\a(s)
\ad(s)}$).  Furthermore, its highest spin component (which is the completely symmetric piece of 
the $\theta\, \bar{\theta}$ term, $h_{\a(s+1)\ad(s+1)}$) must propagate on-shell. But for that to 
happen, it must have mass dimensions one ($[h]=1$) and according to the Fronsdal action of 
massless integer spins (which must be the bosonic piece of our theory) it also needs to be real. 
Therefore our theory must be constructed in terms of a real bosonic superfield $H_{\a(s)\ad(s)}$ 
with zero mass dimensions ($[H]=0$). The most general action that can be written for such an 
object has the form:
\be
\eqalign{
S=\int d^8z\Big\{
&~c_1 H^{\a(s)\ad(s)}\D^{\g}\Dd^2\D_{\g}H_{\a(s)\ad(s)}~+~
c_2 H^{\a(s)\ad(s)}\Box H_{\a(s)\ad(s)}\cr
&+c_3 H^{\a(s)\ad(s)}\pa_{\a_s\ad_s}\pa^{\g\gd}H_{\g\a(s-1)\gd\ad(s-1)}\cr
&+c_4 H^{\a(s)\ad(s)}[\D_{\a_s},\Dd_{\ad_s}][\D^{\g},\Dd^{\gd}]H_{\g\a(s-1)
\gd\ad(s-1)}\Big\}   ~~~. {~~~~~~~~~~~}
} 
\label{equ01}
\ee
In writing this action, we have also made an assumption that parity violating terms should
be excluded.  If this assumption is not used then an additional term of the form
\be
\eqalign{
S_{P-violation}=\int d^8z\Big\{
H^{\a(s)\ad(s)}\pa_{\a_s\ad_s} [\D^{\g},\Dd^{\gd}]H_{\g\a(s-1)
\gd\ad(s-1)}\Big\}   ~~~.
} 
\label{equ02}
\ee
may be considered\footnote{In principle we could repeat the whole analysis including this term and
show that it's coefficient \newline $~\,~~\,~~$ will vanish. But just knowing that the final results are
Frondal's actions for bosons and fermions
\newline $~\,~\,~~~$and they preserves parity, allows us to set this term to zero from the very beginning.}.

The massless property of the theory suggest there must be an underlying gauge 
symmetry.  This symmetry, of course, must respect the highest superspin projection 
operator.  Taking this into account there is only one option.
The gauge transformation of $H_{\a(s)\ad(s)}$ must be of the form:
\be
\delta H_{\a(s)\ad(s)}=\frac{1}{s!}\Dd_{(\ad_s}L_{\a(s)\ad(s-1))}-\frac{1}{s!}\D_{(\a_s}
\bar{L}_{\a(s-1))\ad(s)}  ~~~,
\label{equ03}
\ee
written in terms of some complex gauge parameter superfield $L_{\a(s)\ad(s-1))}$.

The change of the above action under this transformation is :
$$
\eqalign{
\delta S=\int d^8z \Bigg\{
&~\[-2c_1+2c_2+\frac{2}{s}c_3+2\frac{2s+1}{s}c_4\]H^{\a(s)\ad(s)}\Dd_{\ad_s}
\D^2\Dd^2L_{\a(s)\ad(s-1)}   \cr    
}
$$
\be
\eqalign{ {~~\,~~~~}  {~~~~~~~} {~~~~~~~}
&+2c_2 H^{\a(s)\ad(s)}\Dd^2\D^2\Dd_{\ad_s}L_{\a(s)\ad(s-1)}\cr
&+\[-\frac{2}{s}c_3+2\frac{2s+1}{s}c_4\]H^{\a(s)\ad(s)}\D_{\a_s}\Dd^2\D^{\g}
\Dd_{\ad_s}L_{\g\a(s-1)\ad(s-1)}\cr
&+\[2c_3-2c_4\]H^{\a(s)\ad(s)}\Dd_{\ad_s}\D_{\a_s}\Dd^2\D^{\g}L_{\g\a(s-1)
\ad(s-1)}\cr
&-\[\frac{s-1}{s}\]\[2c_3-2c_4\]H^{\a(s)\ad(s)}\Dd_{\ad_s}\D_{\a_s}\Dd_{\ad_{
s-1}}\D^{\g}\Dd^{\gd}L_{\g\a(s-1)\gd\ad(s-2)}\cr
&+c.c.\Bigg\}    ~~~.
}  \label{equ04}
\ee
It is obvious that the above action is not invariant under the proposed gauge transformation.  There
are two ways around this.  One way is to impose differential constraints
(using either D or $\Bar \D$) on the gauge parameter superfield $L_{\a(s)\ad(s-1))}$.
In general this procedure leads to the `ghost-for-ghost' phenomenon in a quantum
theory \cite{Siegel:1980}.  We wish to avoid this. 

The other way, is to introduce a set of compensators. 
In order to keep the propagationing degrees of freedom down to the minimal number and
in order to have on-shell an irreducible representation of the Super-Poincare group, we 
need exactly one `propagating' compensator\footnote{This is a superfield of mass 
dimensions 0 or 1/2.} and some  arbitrary number of auxiliary compensators\footnote{
These are superfields of mass dimensions 1}.

~The propagating compensator must satisfy several constraints. It must provide the 
extra degrees of freedom in order to complete the irreducible representation and the 
rest of its components must vanish on shell. In principle there are two options, it can 
be either bosonic or fermionic. In the first case, the gauge transformation of a bosonic
compensator has to be\footnote{This is fixed just by considering the index structure and mass 
dimensions.} of the form $\D^{\a_s}L_{\a(s)\ad(s-1)}+\Dd^{\ad_s}\bar{L}_{\a(s-1)\ad(s)}$. 
But with a transformation like that we can not gauge away all the degrees of freedom, 
besides the ones needed for the irreducible representation. So this option can not 
lead to the desired result. Therefore the propagating compensator must be a fermionic 
superfield 
$\U_{\a(s)\ad(s-1)}$. This is in accord with the feature of all previous studied
theories, where the statistics of the main superfield and the compensator opposite to
one another.

Also the gauge transformation of the fermionic compensator $\U$ must be such that, 
it satisfies the following:

\begin{itemize}

\item the component $\U^{(1,0)S}_{\a(s+1)\ad(s-1)}$ must be gauged away. This 
can be done if
\be
\D_{(\a_{s+1}}\U_{\a(s))\ad(s-1)}|\sim \text{~some component of the gauge 
parameter (algebraicly)}\nonumber
\ee
 
 \item the component $\U^{(0,1)S}_{\a(s)\ad(s)}$ must be gauged away. This can 
 be done if
 \be
 \Dd_{(\ad_s}\U_{\a(s)\ad(s-1))}|\sim \text{~some component of the gauge parameter 
 (algebraicly)}\nonumber
 \ee
 
 \item the component $\U^{(0,1)A}_{\a(s)\ad(s-2)}$ must be gauged away. This can 
 be done if
\be
\Dd^{\gd}\U_{\a(s)\gd\ad(s-2)}|\sim \text{~some component of the gauge parameter 
(algebraicly)}\nonumber
\ee

\item the component $\U^{(1,1)(A,A)}_{\a(s-1)\ad(s-2)}$ must propagate on shell 
and have a specific gauge transformation. This can be done if
\be
[\D^{\g},\Dd^{\gd}]\U_{\g\a(s-1)\gd\ad(s-2)}|\sim \pa^{\g\gd}\text{~some component 
of the gauge parameter}  \nonumber
\ee
 \end{itemize}
 
The above constraints and equation (\ref{equ04}) will be our guideline.  Based on equation 
(\ref{equ04}), we must find all possible ways that we can introduce a fermionic compensator 
with mass dimensions 1/2 and with the specific index structure $\U_{\a(s)\ad(s-1)}$, which 
has a gauge transformation that satisfies all the above constraints. This can be done only 
by two ways.
 
 ~Since $L_{\a(s)\ad(s-1)}$ and $\bar{L}_{\a(s-1)\ad(s)}$ are the only gauge parameters 
 available, the gauge transformation of $\U$ must include at least one of them. Their mass 
 dimensions are -1/2 ($[L]=[\bar{L}]=-1/2$), so we need 2 $\D(\Dd)$'s to build something 
 with mass dimensions 1/2. Therefore, the transformation of $\U_{\a(s)\ad(s-1)}$ must include at 
 least one of the following terms.
 \be
 \eqalign{
 &A) {~~~} \Dd^2L_{\a(s)\ad(s-1)}   ~~~, \cr\nonumber
 &B)  {~~~} \frac{1}{s!}\Dd^{\ad_s}\D_{(\a_s}\bar{L}_{\a(s-1))\ad(s)} ~~~,  \cr
 &C)  {~~~}\D^2L_{\a(s)\ad(s-1)} ~~~, \cr
 &D)  {~~~} \frac{1}{s!}\D_{(\a_s}\Dd^{\ad_s}\bar{L}_{\a(s-1)\ad(s)} ~~~.
 }
 \ee
 
 In order, these types of possible transformations to have a hope to give 
 something desirable, they need to be completed appropriately so the above constraints
 are satisfied. The minimal way to do that is the following:
 \be
 \eqalign{
 &A) {~~~}  \Dd^2L_{\a(s)\ad(s-1)}+\D^{\a_{s+1}}\Lambda_{\a(s+1)\ad(s)} ~~~, \cr\nonumber
 &B) {~~~} \frac{1}{s!}\Dd^{\ad_s}\D_{(\a_s}\bar{L}_{\a(s-1))\ad(s)}+\D^{\a_{s+1}}
 \Lambda_{\a(s+1)\ad(s)} ~~~, \cr
 &C) {~~~}\D^2L_{\a(s)\ad(s-1)}+\Dd^{\ad_s}V_{\a(s)\ad(s)} ~~~.
 }
 \ee
 the last case was eliminated through this last requirement.
 
~Now it is really straightforward to check if any of the above transformations can be 
used together with (\ref{equ04}) in order to introduce the fermionic compensator. Just 
by observing (\ref{equ04}) we see that case C can not happen and we are left with two 
possibilities. Case (A) can arise from the first term of (\ref{equ04}) and Case (B) can arise 
from the third term.  Next we will study this two cases.

\section{The Higher Superspin KPS-Series}

For  Case (A) consider, we impose
\be
c_2=c_3=c_4=0 ~~~,
\label{equ05}
\ee
so then equation (\ref{equ04} )becomes:
\be
\eqalign{
\delta S&=\int d^8z\Big\{-2c_1H^{\a(s)\ad(s)}\Dd_{\ad_s}\D^2\Dd^2L_{\a(s)\ad(s-1)}+c.c.\Big\}
\cr
&=\int d^8z\Big\{-2c_1H^{\a(s)\ad(s)}\Dd_{\ad_s}\D^2\[\Dd^2L_{\a(s)\ad(s-1)}+\D^{\a_{s+1}}
\Lambda_{\a(s+1)\ad(s-1)}\]+c.c.\Big\}  ~~~.
}
\label{equ06}
\ee

At this point we can introduce a fermionic compensator $\U_{\a(s)\ad(s-1)}$ with the following 
gauge transformation:
\be
\delta\U_{\a(s)\ad(s-1)}=\Dd^2L_{\a(s)\ad(s-1)}+\D^{\a_{s+1}}\Lambda_{\a(s+1)\ad(s)}
~~~,
\label{equ07}
\ee
and in order to construct a fully invariant action and give to the compensator some dynamics 
we have to add to the initial action some more terms

\begin{itemize}
\item Add a counter term, which cancels the chang of the initial action:
\be
S_c=\int d^8z~2c_1H^{\a(s)\ad(s)}\left(\Dd_{\ad_s}\D^2\U_{\a(s)\ad(s-1)}-\D_{\a_s}\Dd^2
\bar{\U}_{\a(s-1)\ad(s)}\right)
\label{equ08}
\ee

\item Add a kinetic energy term for the compensator (the most general free action quadratic to 
$\U_{\a(s)\ad(s-1)}$)
\be
\eqalign{
S_{k.e}=\int d^8z\Big\{
&~h_1\U^{\a(s)\ad(s-1)}\D^2\U_{\a(s)\ad(s-1)}+c.c.\cr
&+h_2\U^{\a(s)\ad(s-1)}\Dd^2\U_{\a(s)\ad(s-1)}+c.c.\cr
&+h_3\U^{\a(s)\ad(s-1)}\Dd^{\ad_s}\D_{\a_s}\bar{\U}_{\a(s-1)\ad(s)}\cr
&+h_4\U^{\a(s)\ad(s-1)}\D_{a_s}\Dd^{\ad_s}\bar{\U}_{\a(s-1)\ad(s)}\Big\}
}
\label{equ09}
\ee
\end{itemize}

The full action is thus given by
\be
\eqalign{  {~~~~~~~~~}
S_{Full}=\int d^8z\Big\{
&~c_1 H^{\a(s)\ad(s)}\D^{\g}\Dd^2\D_{\g}H_{\a(s)\ad(s)}\cr
&+2c_1H^{\a(s)\ad(s)}\left(\Dd_{\ad_s}\D^2\U_{\a(s)\ad(s-1)}-\D_{\a_s}\Dd^2\bar{
\U}_{\a(s-1)\ad(s)}\right)\cr
&+h_1\U^{\a(s)\ad(s-1)}\D^2\U_{\a(s)\ad(s-1)}+c.c.\cr
&+h_2\U^{\a(s)\ad(s-1)}\Dd^2\U_{\a(s)\ad(s-1)}+c.c.\cr
&+h_3\U^{\a(s)\ad(s-1)}\Dd^{\ad_s}\D_{\a_s}\bar{\U}_{\a(s-1)\ad(s)}\cr
&+h_4\U^{\a(s)\ad(s-1)}\D_{a_s}\Dd^{\ad_s}\bar{\U}_{\a(s-1)\ad(s)}\Big\}
~~~. }
\label{equ10}
\ee

Now we can define the superfields ${\bm  {{\cal G}}}_{\a(s)\ad(s)}$ and ${\bm{{\cal T}}}_{\a(s)
\ad(s-1)}$ as the variations of the full action with respect to the superfields $H_{\a(s)
\ad(s)}$ and $\U_{\a(s)\ad(s-1)}$.  These variations yield respectively,
\be
\eqalign{
{\bm{{\cal G}}}_{\a(s)\ad(s)}=&~2c_1\D^{\g}\Dd^2\D_{\g}H_{\a(s)\ad(s)}
\,+\, \frac{2c_1}{s!}\left(\Dd_{(\ad_s}\D^2\U_{\a(s)\ad(s-1))}-\D_{(\a_s}\Dd^2\bar{
\U}_{\a(s-1))\ad(s)}\right)  ~~~,
}
\label{equ11}
\ee
and
\be
\eqalign{ {~~~}
{\bm{{\cal T}}}_{\a(s)\ad(s-1)}=&~2c_1\D^2\Dd^{\ad_s}H_{\a(s)\ad(s)}  
\,+\, 2h_1\D^2\U_{\a(s)\ad(s-1)} \,+\, 2h_2\Dd^2\U_{\a(s)\ad(s-1)}\cr
&+\, \frac{h_3}{s!}\Dd^{\ad_s}\D_{(\a_s}\bar{\U}_{\a(s-1))\ad(s)}
\,+\, \frac{h_4}{s!} \D_{(a_s}\Dd^{\ad_s}\bar{\U}_{\a(s-1))\ad(s)} ~~~.
}
\label{equ12}
\ee

The invariance of the full action under the above gauge transformations, forces a set of 
constraints that must be satisfied. These are the Bianchi Identities which are going 
to determine all the free parameters.
\be
\eqalign{
&\Dd^{\ad_s}{\bm{{\cal G}}}_{\a(s)\ad(s)}+\Dd^2{\bm{{\cal T}}}_{\a(s)\ad(s-1)}=0 ~~~,  \cr
&  \D_{(\a_{s+1}}{\bm{{\cal T}}}_{\a(s))\ad(s-1)}=0 ~~~.
}
\label{equ13}
\ee
The solution of the first one is:
\be
\eqalign{
&h_1=-\frac{s+1}{s}c_1   ~~~, ~~~
h_4=2c_1  ~~~,
}
\label{equ14}
\ee
and the solution of the second one is:
\be
\eqalign{
&h_2=0  ~~~,~~~
h_3=0  ~~~.
} 
\label{equ15}
\ee

Therefore the final action is:
\be
\eqalign{
S=\int d^8z\Big\{
&~c_1H^{\a(s)\ad(s)}\D^{\g}\Dd^2\D_{\gd}H_{\a(s)\ad(s)}\cr
&+2c_1H^{\a(s)\ad(s)}\left(\Dd_{\ad_s}\D^2\U_{\a(s)\ad(s-1)}
-\D_{\a_s}\Dd^2\bar{\U}_{\a(s-1)\ad(s)}\right)\cr
&-\[\frac{s+1}{s}\]c_1\U^{\a(s)\ad(s-1)}\D^2\U_{\a(s)\ad(s-1)}+c.c.\cr
&+2c_1\U^{\a(s)\ad(s-1)}\D_{a_s}\Dd^{\ad_s}\bar{\U}_{\a(s-1)\ad(s)}\Big\}   ~~~,
}
\label{equ16}
\ee
and it is invariant under the gauge transformations
\be
\eqalign{
&\delta H_{\a(s)\ad(s)}=\frac{1}{s!}\Dd_{(\ad_s}L_{\a(s)\ad(s-1))}-\frac{1}{s!}\D_{(\a_s}
\bar{L}_{\a(s-1))\ad(s)}    ~~~~,\cr
&\delta\U_{\a(s)\ad(s-1)}=\Dd^2L_{\a(s)\ad(s-1)}+\D^{\a_{s+1}}\Lambda_{\a(s+1)\ad(s)}
~~~.
}
\label{equ17}
\ee
This theory is equivalent to that developed by Kuzenko, Postnikov, and Sibiryakov (KPS) \cite{
Kuzenko:1993jp}, once one solves the constraints that appear in their description (as done in \cite{Gates:2010td}).
Therefore without any further examination we can conclude that this action, with that set of transformations 
describes a massless half odd superspin ($Y=s+1/2$).

\section{The Higher Superspin B-Series}

For Case (B) we impose
\be
\eqalign{
c_2&=0   ~~~, \cr
-2c_1+\frac{2}{s}c_3+2\frac{2s+1}{s}c_4&=0~~\Rightarrow c_1=\frac{1}{s}\left[c_3+(2s+1)c_4\right]
~~~.
}
\label{equ18}
\ee
so that equation (\ref{equ04}) becomes:
$$
\eqalign{
\delta S=\int d^8z \, {\Big \{ }
&\[\frac{2}{s}\](c_3-(2s+1)c_4)H^{\a(s)\ad(s)}\Dd_{\ad_s}\D^2\Big[\frac{1}{s!}\Dd^{\gd}\D_{
(\a_s}\bar{L}_{\a(s-1))\gd\ad(s-1)}\Big]\cr
+&\[\frac{2}{s}\](c_3-(2s+1)c_4)H^{\a(s)\ad(s)}\Dd_{\ad_s}\D^2\Big[\D^{\a_{s+1}}\Lambda_{
\a(s+1)\ad(s-1)}\Big]\cr
+& c.c.\cr
-&2(c_3-c_4)H^{\a(s)\ad(s)}[\D_{\a_s},\Dd_{\ad_s}]\Big[\Dd^2\D^{\g}L_{\g\a(s-1)\ad(s-1)}
\Big]\cr
}
$$

\be
\eqalign{
{~~~~~~~~~~~~~~~~~~~~} -&2(c_3-c_4)H^{\a(s)\ad(s)}[\D_{\a_s},\Dd_{\ad_s}]\Big[\D^2\Dd^{\gd}\bar{L}_{\a(s-1)
\gd\ad(s-1)}\Big]\cr
-&2(c_3-c_4)H^{\a(s)\ad(s)}[\D_{\a_s},\Dd_{\ad_s}]\Big[-\frac{s-1}{s!}\Dd_{(\ad_{s-1}}
\D^{\g}\Dd^{\gd}L_{\g\a(s-1)\gd\ad(s-2))}\Big]\cr
-&2(c_3-c_4)H^{\a(s)\ad(s)}[\D_{\a_s},\Dd_{\ad_s}]\Big[-\frac{s-1}{s!}\D_{(\a_{s-1}}
\Dd^{\gd}\D^{\g}\bar{L}_{\g\a(s-2))\gd\ad(s-1)}\Big]  {\Big \} }~~~.
}
\label{equ19}
\ee
 
We introduce two compensators: \\
1) A fermionic propagating compensator $\U_{\a(s)\ad(s-1)}$ with mass dimensions 1/2 \newline 
$~~~~$
and the following gauge transformation
 \be
 \delta\U_{\a(s)\ad(s-1)}=\frac{1}{s!}\Dd^{\gd}\D_{(\a_s}\bar{L}_{\a(s-1))\gd\ad(s-1)}
 +\D^{\a_{s+1}}\Lambda_{\a(s+1)\ad(s-1)} ~~~.
\label{equ20}
 \ee
 2) A real auxiliary bosonic compensator $B_{\a(s-1)\ad(s-1)}$, with mass dimensions 1 
 \newline $~~~~$
 which 
 transforms as
 \be
 \eqalign{
 \delta B_{\a(s-1)\ad(s-1)}=&~\Dd^2\D^{\a_s}L_{\a(s)\ad(s-1)}+\D^2\Dd^{\ad_s}
 \bar{L}_{\a(s-1)\ad(s)}\cr
 &-\[\frac{s-1}{s!}\]\Dd_{(\ad_{s-1}}\D^{\g}\Dd^{\gd}L_{\g\a(s-1)\gd\ad(s-2))}\cr
 &-\[\frac{s-1}{s!}\]\D_{(\a_{s-1}}\Dd^{\gd}\D^{\g}\bar{L}_{\g\a(s-2))\gd\ad(s-1)} 
 ~~~.}
\label{equ21}
\ee
 
 To create an invariant action and give dynamics to the compensators we have to add the 
 following terms:
 \begin{itemize}
 \item A counter term which will cancel the change of the initial action
 \be
 \eqalign{
 S_c=\int d^8z\Big\{&-\[\frac{2}{s}\](c_3-(2s+1)c_4)H^{\a(s)\ad(s)}\Dd_{\ad_s}\D^2\U_{
 \a(s)\ad(s-1)}+c.c.\cr
&+2(c_3-c_4)H^{\a(s)\ad(s)}[\D_{\a_s},\Dd_{\ad_s}]B_{\a(s-1)\ad(s-1)}\Big\}
~~~, }
\label{equ22}
\ee

\item A kinetic energy term for both the compensators (the most general action for $\U$ and $B$)
\be
\eqalign{
S_{k.e}=\int d^8z\Big\{
&~eB^{\a(s-1)\ad(s-1)}B_{\a(s-1)\ad(s-1)}\cr
&+h_1\U^{\a(s)\ad(s-1)}\D^2\U_{\a(s)\ad(s-1)}+c.c.\cr
&+h_2\U^{\a(s)\ad(s-1)}\Dd^2\U_{\a(s)\ad(s-1)}+c.c.\cr
&+h_3\U^{\a(s)\ad(s-1)}\Dd^{\ad_s}\D_{\a_s}\bar{\U}_{\a(s-1)\ad(s)}\cr
&+h_4\U^{\a(s)\ad(s-1)}\D_{a_s}\Dd^{\ad_s}\bar{\U}_{\a(s-1)\ad(s)}\Big\}
~~~,}
\label{equ23}
\ee

\item An interaction term among compensators (as in principle, such  a term can exist)
\be
S_{int}=\int d^8z \Big\{bB^{\a(s-1)\ad(s-1)}\[D^{\a_s}\U_{\a(s)\ad(s-1)}+\Dd^{\ad_s}\bar{\U}_{
\a(s-1)\ad(s)}\]\Big\}  ~~~.
\label{equ24}
\ee
\end{itemize}

Therefore the full action is
\be
\eqalign{
S=\int d^8z\Big\{
&~\[\frac{1}{s}\](c_3+(2s+1)c_4) H^{\a(s)\ad(s)}\D^{\g}\Dd^2\D_{\g}H_{\a(s)\ad(s)}\cr
&+c_3 H^{\a(s)\ad(s)}\pa_{\a_s\ad_s}\pa^{\g\gd}H_{\g\a(s-1)\gd\ad(s-1)}\cr
&+c_4 H^{\a(s)\ad(s)}[\D_{\a_s},\Dd_{\ad_s}][\D^{\g},\Dd^{\gd}]H_{\g\a(s-1)\gd\ad(s-1)}\cr
&-\[\frac{2}{s}\](c_3-(2s+1)c_4)H^{\a(s)\ad(s)}\Dd_{\ad_s}\D^2\U_{\a(s)\ad(s-1)}\cr
&+\[\frac{2}{s}\](c_3-(2s+1)c_4)H^{\a(s)\ad(s)}\D_{\a_s}\Dd^2\bar{\U}_{\a(s-1)\ad(s)}\cr
&+2(c_3-c_4)H^{\a(s)\ad(s)}[\D_{\a_s},\Dd_{\ad_s}]B_{\a(s-1)\ad(s-1)}\cr
&+eB^{\a(s-1)\ad(s-1)}B_{\a(s-1)\ad(s-1)}\cr
&+h_1\U^{\a(s)\ad(s-1)}\D^2\U_{\a(s)\ad(s-1)}+c.c.\cr
&+h_2\U^{\a(s)\ad(s-1)}\Dd^2\U_{\a(s)\ad(s-1)}+c.c.\cr
&+h_3\U^{\a(s)\ad(s-1)}\Dd^{\ad_s}\D_{\a_s}\bar{\U}_{\a(s-1)\ad(s)}\cr
&+h_4\U^{\a(s)\ad(s-1)}\D_{a_s}\Dd^{\ad_s}\bar{\U}_{\a(s-1)\ad(s)}\cr
&+bB^{\a(s-1)\ad(s-1)}\left(D^{\a_s}\U_{\a(s)\ad(s-1)}+\Dd^{\ad_s}\bar{\U}_{\a(s-1)\ad(s)}
\right)\Big\}
~~~.
} \label{equ25}
\ee

The invariance of this action under the corresponding gauge transformations is guaranteed 
by the satisfaction of the following two Bianchi identities
\be
\eqalign{  {~~~~}
0=&~\D^{\a_s}{\bm{{\cal G}}}_{\a(s)\ad(s)}-\frac{1}{s!}\D^{\a_s}\Dd_{(\ad_s}{\bm{{\cal T}}}_{
\a(s)\ad(s-1))}
~~~, \cr
&+\frac{1}{s!}\Dd_{(\ad_s}\D^2{\bm{{\cal Y}}}_{\a(s-1)\ad(s-1))}-\frac{s-1}{s!s!}\D_{(\a_{s-1}}
\Dd_{(\ad_s}\D^{\g}{\bm{{\cal Y}}}_{\g\a(s-2))\ad(s-1))}=0 ~~~, \cr
0=&~ \D_{(\a_{s+1}}{\bm{{\cal T}}}_{\a(s))\ad(s-1)} ~~~, 
} \label{equ26}
\ee
where ${\bm{{\cal G}}}_{\a(s)\ad(s)},~{\bm{{\cal T}}}_{\a(s)\ad(s-1)},~{\bm{{\cal Y}}}_{\a(s-1)
\ad(s-1)}$ are the variations of the action with respect the corresponding superfields $H_{
\a(s)\ad(s)},~\U_{\a(s)\ad(s-1)}$, and $B_{\a(s-1)\ad(s-1)}$.

The solution of the first bianchi identity gives:
$$
\eqalign{
h_1 &=-\[\frac{1}{s}\]\left(c_3-(2s+1)c_4\right)  ~~~, ~~~
h_3 =0  ~~~, \cr
h_4 &=\[\frac{2(s+1)}{s^2}\]\left(c_3-(2s+1)c_4\right) ~~~, \cr
}
$$
\be
\eqalign{
e &=\frac{1}{2}b-(c_3-c_4)  ~~~, \cr
b&=-h_4=-\[\frac{2(s+1)}{s^2}\]\left(c_3-(2s+1)c_4\right) ~~~,  \cr
b&=\[\frac{2(2s+1)}{s}\]\left(c_3-c_4\right) ~~~.
}
\label{equ27}
\ee
the last two equations will give a relationship among $c_3$ and $c_4$
\be
c_4=\[\frac{2s^2+2s+1}{(2s+1)^2}\]c_3  ~~~.
\label{equ28}
\ee
The second Binachi identity has as a solution:
\be
\eqalign{
&h_2=0  ~~,~~~
h_3=0 ~~~.
}
\label{equ29}
\ee

So the full action takes the form
\be
\eqalign{
S=\int d^8z\Big\{
&~\[\frac{2(s+1)^2}{s(2s+1)}\]c_3H^{\a(s)\ad(s)}\D^{\g}\Dd^2\D_{\g}H_{\a(s)\ad(s)}\cr
&+c_3 H^{\a(s)\ad(s)}\pa_{\a_s\ad_s}\pa^{\g\gd}H_{\g\a(s-1)\gd\ad(s-1)}\cr
&+ \[\frac{2s^2+2s+1}{(2s+1)^2}\]c_3H^{\a(s)\ad(s)}[\D_{\a_s},\Dd_{\ad_s}][\D^{\g},
\Dd^{\gd}]H_{\g\a(s-1)\gd\ad(s-1)}\cr
&+\[\frac{4s}{2s+1}\]c_3H^{\a(s)\ad(s)}\Dd_{\ad_s}\D^2\U_{\a(s)\ad(s-1)}\cr
&-\[\frac{4s}{2s+1}\]c_3H^{\a(s)\ad(s)}\D_{\a_s}\Dd^2\bar{\U}_{\a(s-1)\ad(s)}\cr
&+\[\frac{4s(s+1)}{(2s+1)^2}\]c_3H^{\a(s)\ad(s)}[\D_{\a_s},\Dd_{\ad_s}]B_{\a(s-1)\ad(s-1)}\cr
&+\[\frac{2(s+1)^2}{(2s+1)^2}\]c_3B^{\a(s-1)\ad(s-1)}B_{\a(s-1)\ad(s-1)}\cr
&+\[\frac{2s}{2s+1}\]c_3\U^{\a(s)\ad(s-1)}\D^2\U_{\a(s)\ad(s-1)}+c.c.\cr
&-\[\frac{4(s+1)}{2s+1}\]c_3\U^{\a(s)\ad(s-1)}\D_{a_s}\Dd^{\ad_s}\bar{\U}_{\a(s-1)
\ad(s)}\cr
&+\[\frac{4(s+1)}{2s+1}\]c_3B^{\a(s-1)\ad(s-1)}\left(D^{\a_s}\U_{\a(s)\ad(s-1)}+
\Dd^{\ad_s}\bar{\U}_{\a(s-1)\ad(s)}\right)\Big\}
~~~. }
\label{equ30}
\ee

At this point we can use the equation of motion of the auxiliary superfield $B_{\a(s-1)\ad(s-1)}$
\be
\eqalign{
&{\bm{{\cal Y}}}_{\a(s-1)\ad(s-1)}=0\Rightarrow\cr
&B_{\a(s-1)\ad(s-1)}=-\[\frac{s}{s+1}\][\D^{\a_s},\Dd^{\ad_s}]H_{\a(s)\ad(s)}\cr
&~~~~~~~~~~~~~~~~~~~-\[\frac{2s+1}{s+1}\]\left(D^{\a_s}\U_{\a(s)\ad(s-1)}+\Dd^{\ad_s}
\bar{\U}_{\a(s-1)\ad(s)}\right) ~~~,
}
\label{equ31}
\ee
in order to integrate it out and simplify the action.  So our final action is:
\be
\eqalign{
S=\int d^8z\Big\{
&~\[\frac{2(s+1)^2}{s(2s+1)}\]c_3H^{\a(s)\ad(s)}\D^{\g}\Dd^2\D_{\g}H_{\a(s)\ad(s)}\cr
&+c_3 H^{\a(s)\ad(s)}\pa_{\a_s\ad_s}\pa^{\g\gd}H_{\g\a(s-1)\gd\ad(s-1)}\cr
&+ \[\frac{1}{(2s+1)}\]c_3H^{\a(s)\ad(s)}[\D_{\a_s},\Dd_{\ad_s}][\D^{\g},\Dd^{\gd}]H_{\g\a(s-1)
\gd\ad(s-1)}\cr
&-\[\frac{4s}{2s+1}\]c_3H^{\a(s)\ad(s)}\D_{\a_s}\Dd_{\ad_s}\D^{\g}\U_{\g\a(s-1)\ad(s-1)}\cr
&+\[\frac{4s}{2s+1}\]c_3H^{\a(s)\ad(s)}\Dd_{\ad_s}\D_{\a_s}\Dd^{\gd}\bar{\U}_{\a(s-1)
\gd\ad(s-1)}\cr
&-\[\frac{2(s+1)}{2s+1}\]c_3\U^{\a(s)\ad(s-1)}\D^2\U_{\a(s)\ad(s-1)}+c.c.\cr
&+\[\frac{4s}{2s+1}\]c_3\U^{\a(s)\ad(s-1)}\D_{a_s}\Dd^{\ad_s}\bar{\U}_{\a(s-1)\ad(s)}\Big\}
~~~, }
\label{equ32}
\ee
and it is invariant under the following gauge transformations
\be 
\eqalign{
&\delta H_{\a(s)\ad(s)}=\frac{1}{s!}\Dd_{(\ad_s}L_{\a(s)\ad(s-1))}-\frac{1}{s!}\D_{(\a_s}\bar{L
}_{\a(s-1))\ad(s)}\cr
&\delta\U_{\a(s)\ad(s-1)}=\frac{1}{s!}\Dd^{\gd}\D_{(\a_s}\bar{L}_{\a(s-1))\gd\ad(s-1)}+\D^{
\a_{s+1}}\Lambda_{\a(s+1)\ad(s-1)}
~~~.}
\label{equ33}
\ee
From this action we can calculate the following superfields  
$$
\eqalign{
{\bm{{\cal G}}}_{\a(s)\ad(s)}=&~\[\frac{4(s+1)^2}{s(2s+1)}\]c_3\D^{\g}\Dd^2\D_{\g}H_{\a(s)\ad(s)}\cr
&+\frac{2c_3}{s!s!}\pa_{(\a_s(\ad_s}\pa^{\g\gd}H_{\g\a(s-1))\gd\ad(s-1))}\cr
&+ \[\frac{2}{(2s+1)}\]\frac{c_3}{s!s!}[\D_{(\a_s},\Dd_{(\ad_s}][\D^{\g},\Dd^{\gd}]H_{\g\a(s-1))\gd
\ad(s-1))}\cr
&-\[\frac{4s}{2s+1}\]\frac{c_3}{s!s!}\D_{(\a_s}\Dd_{(\ad_s}\D^{\g}\U_{\g\a(s-1))\ad(s-1))}\cr
&+\[\frac{4s}{2s+1}\]\frac{c_3}{s!s!}\Dd_{(\ad_s}\D_{(\a_s}\Dd^{\gd}\bar{\U}_{\a(s-1))\gd\ad(
s-1))}\cr
~~~,}
$$
\be
\eqalign{
{\bm{{\cal T}}}_{\a(s)\ad(s-1)}=&-\[\frac{4s}{2s+1}\]\frac{c_3}{s!}\D_{(\a_s}\Dd^{\gd}\D^{\g}H_{
\g\a(s-1))\ad(s-1)}\cr
&-\[\frac{4(s+1)}{2s+1}\]c_3\D^2\U_{\a(s)\ad(s-1)}\cr
&+\[\frac{4s}{2s+1}\]\frac{c_3}{s!}\D_{(a_s}\Dd^{\ad_s}\bar{\U}_{\a(s-1))\ad(s)}
~~~,}
\label{equ34}
\ee
 which satisfhy the bianchi identities for this action
 \be
\eqalign{
0=&~\D^{\a_s}{\bm{{\cal G}}}_{\a(s)\ad(s)}-\frac{1}{s!}\D^{\a_s}\Dd_{(\ad_s}{\bm{{\cal T}}}_{\a(s)
\ad(s-1))}\cr
0=&~  \D_{(\a_{s+1}}{\bm{{\cal T}}}_{\a(s))\ad(s-1)}
~~~.}
\label{equ35}
\ee
 
 It is also straightforward to prove that they satisfy another identity
 \be
 \eqalign{
 \D^{\a_{2s+1}}{\bm{\cal W}}_{\a(2s+1)}=&~\[\frac{s(2s+1)}{4(s+1)^2}\]\frac{1}{c_3}\pa_{(\a_{2s}}
 {}^{\ad_s}\dots\pa_{(\a_{s+1}}{}^{\ad_1}{\bm{{\cal G}}}_{\a(s))\ad(s)}\cr
 &+i\[\frac{s^2}{4(s+1)^2}\]\frac{1}{c_3}\D_{(\a_{2s}}\Dd^2\pa_{(\a_{2s-1}}{}^{\ad_{s-1}}\dots\pa_{
 (\a_{s+1}}{}^{\ad_1}{\bm{{\cal T}}}_{\a(s))\ad(s-1)}\cr
 &+\[\frac{s^2}{4(s+1)^2}\]\frac{1}{c_3}\D_{(\a_{2s}}\pa_{(\a_{2s-1}}{}^{\ad_{s}}\dots\pa_{(\a_{s}}
 {}^{\ad_1}\bar{{\bm {{}\cal T}}}_{\a(s-1))\ad(s)}
 ~~~, }
 \label{equ36}
 \ee
 with
 \be
 {\bm{\cal W}}_{\a(2s+1)}=\Dd^2\D_{(\a_{2s+1}}\pa_{(\a_{2s}}{}^{\ad_{s}}\dots\pa_{(\a_{s+1}}
 {}^{\ad_1}H_{\a(s))\ad(s)} ~~~.
\label{equ37}
 \ee
 That means that on-shell (${\bm{{\cal T}}}={\bm{{\cal G}}}=0$) the object ${\bm{\cal 
 W}}_{\a(2s+1)}$ satisfies the 
 equations
 \be
 \eqalign{
 &\D^{\a_{2s+1}}{\bm{\cal W}}_{\a(2s+1)}=0  ~~~, ~~~
 \Dd_{\ad}{\bm{\cal W}}_{\a(2s+1)}=0  ~~~,
 }
 \label{equ38}
 \ee
therefore it describes a massless half odd superspin.  Now we know that this theory, 
on-shell has an irreducible representation propagating. The last thing we need to check 
is whether these are the only degrees of freedom propagating or if there are more. The 
easiest way to do that is to go to components notation and calculate the action in the 
Wess-Zumino gauge. If the only thing propagating is this half odd supermultiplet, 
the components action must be the Fronsdal action for bosons and fermions respectively.
 
 ~Because of the gauge transformation, we have the freedom to gauge away some 
 of the components. Specifically:\footnote{The definition of symmetric and antisymmetric
 pieces of a field is the following $$\Phi_{\g\a(s-1)}=\Phi^{(S)}_{\g\a(s-1)}+\frac{s-1}{s!}
 C_{\g(\a_{s-1}}\Phi^{(A)}_{\a(s-2))}~~,~~\Phi^{(S)}_{\g\a(s-1)}=\frac{1}{s!}\Phi_{(\g\a(s-1))}
 ~~,~~\Phi^{(A)}_{\a(s-2)}=C^{\g\a_{s-1}}\Phi_{\g\a(s-1)}$$
 Furthermore the notation $\Phi^{(m,n)}$ represents the $\theta^{m}\bar{\theta}^n$ 
 component in the taylor series of the superfield $\Phi$}\\
\begin{tabular}{c c} 
 Bosons: & Fermions: \\
\begin{tabular}{| l | l |}
\hline
Component & Gauged away by\\
\hline
$H^{(0,0)}_{\a(s)\ad(s)}$ & $Re\left[L^{(0,1)(S)}_{\a(s)\ad(s)}\right]$\\
\hline
$H^{(2,0)}_{\a(s)\ad(s)}$ & $L^{(2,1)(S)}_{\a(s)\ad(s)}$\\
\hline
$H^{(1,1)(A,S)}_{\a(s-1)\ad(s+1)}$ & $\bar{L}^{(2,1)(S)}_{\a(s-1)\ad(s+1)}$\\
\hline
$\U^{(1,0)(S)}_{\a(s+1)\ad(s-1)}$ & $\Lambda^{(2,0)}_{\a(s+1)\ad(s-1)}$\\
\hline
$\U^{(1,0)(A)}_{\a(s-1)\ad(s-1)}$ & $\bar{L}^{(2,1)(A)}_{\a(s-1)\ad(s-1)}$\\ 
\hline
$\U^{(0,1)(S)}_{\a(s)\ad(s)}$ & $\Lambda^{(1,1)(A,S)}_{\a(s)\ad(s)}$\\
\hline
$\U^{(0,1)(A)}_{\a(s)\ad(s-2)}$ & $\Lambda^{(1,1)(A,A)}_{\a(s)\ad(s-2)}$\\
\hline
$\U^{(1,2)(S)}_{\a(s+1)\ad(s-1)}$ & $\Lambda^{(2,2)}_{\a(s+1)\ad(s-1)}$\\
\hline
\end{tabular}
&
\begin{tabular}{| l | l |}
\hline
Component & Gauged away by\\
\hline
$H^{(1,0)(S)}_{\a(s+1)\ad(s)}$ & $L^{(1,1)(S,S)}_{\a(s+1)\ad(s)}$\\
\hline
$H^{(1,0)(A)}_{\a(s-1)\ad(s)}$ & $\bar{L}^{(2,0)}_{\a(s-1)\ad(s)}$\\
\hline
$\U^{(0,0)}_{\a(s)\ad(s-1)}$ & $\Lambda^{(1,0)(A)}_{\a(s)\ad(s-1)}$\\
\hline
$\U^{(0,2)}_{\a(s)\ad(s-1)}$ & $\Lambda^{(1,2)(A)}_{\a(s)\ad(s-1)}$\\
\hline
$\U^{(1,1)(S,S)}_{\a(s+1)\ad(s)}$ & $\Lambda^{(2,1)(S)}_{\a(s+1)\ad(s)}$\\
\hline
$\U^{(1,1)(S,A)}_{\a(s+1)\ad(s-2)}$ & $\Lambda^{(2,1)(A)}_{\a(s+1)\ad(s-2)}$\\
\hline
$\U^{(1,1)(A,S)}_{\a(s-1)\ad(s)}$ & $\bar{L}^{(2,2)}_{\a(s-1)\ad(s)}$\\
\hline
\end{tabular}
\\
\end{tabular}

So in the Wess-Zumino gauge for the two superfields are:
\be
\eqalign{
H_{\a(s)\ad(s)}=&~\theta^{\a_{s+1}}\bar{\theta}^{\ad_{s+1}}h_{\a(s+1)\ad(s+1)}-
\frac{s}{s!s!}\theta_{(\a_{s}}\bar{\theta}_{(\ad_{s}}h_{\a(s-1))\ad(s-1))}\cr
&+\frac{1}{\sqrt{2}}\bar{\theta}^2\theta^{\a_{s+1}}\psi_{\a(s+1)\ad(s)}+\frac{1}{\sqrt{2}}
\theta^2\bar{\theta}^{\ad_{s+1}}
\bar{\psi}_{\a(s)\ad(s+1)}\cr
&+\frac{1}{\sqrt{2}s!}\theta^2\bar{\theta}_{(\ad_s}\psi_{\a(s)\ad(s-1)}-\frac{1}{\sqrt{2}s!}
\bar{\theta}^2
\theta_{(\a_s}\bar{\psi}_{\a(s-1))\ad(s)}\cr
&+\theta^2\bar{\theta}^2{A}_{\a(s)\ad(s)} ~~~,
} \label{equ39}
\ee
and
\be
\eqalign{
\U_{\a(s)\ad(s-1)}=&~\theta^2\[\rho_{\a(s)\ad(s-1)}+\frac{1}{\sqrt{2}}\psi_{\a(s)\ad(s-1)}\]\cr
&+\frac{1}{\sqrt{2}s!(s-1)!}\theta_{(\a_s}\bar{\theta}_{(\ad_{s-1}}\psi_{\a(s-1))\ad(s-2))}\cr
&+\theta^2\bar{\theta}^{\ad_s}\[v_{\a(s)\ad(s)}+iw_{\a(s)\ad(s)}-\frac{s}{2s+1}A_{\a(s)
\ad(s)}\right.\cr
&~~~~~~~~~~~~-i\frac{s^2+2s+2}{2(s!)^2}\pa_{(\a_s(\ad_s}h_{\a(s-1))\ad(s-1))}\cr
&~~~~~~~~~~~~+i\left.\frac{s}{2}\pa^{\a_{s+1}\ad_{s+1}}h_{\a(s+1)\ad(s+1)}\]\cr
&+\frac{s-1}{s!}\theta^2\bar{\theta}_{(\ad_{s-1}}\[U_{a(s)\ad(s-2))}+i\frac{s+1}{s!}
\pa_{(\a_s}{}^{\gd}h_{\a(s-1))\gd\ad(s-2))}\]\cr
&+\frac{s}{(s+1)!}\bar{\theta}^2\theta_{(\a_s}\[S_{\a(s-1))\ad(s-1)}+i P_{\a(s-1)\ad(s-1)}\]\cr
&+\theta^2\bar{\theta}^2\[\beta_{\a(s)\ad(s-1)}+\frac{i}{2\sqrt{2}s!(s-1)!}\pa_{(\a_s(\ad_{
s-1}}\psi_{\a(s-1))\ad(s-2))}\right.\cr
&~~~~~~-\frac{i}{\sqrt{2}(s+1)!}\pa_{\a_s}{}^{\ad_s}\bar{\psi}_{\a(s-1))\ad(s)}
+\left.\frac{i}{2}\frac{s}{(s+1)!}\pa_{(\a_s}{}^{\ad_s}\bar{\rho}_{\a(s-1))\ad(s)}\]}\label{equ40}
\ee

From mass dimensions arguments we can tell immediately that the components $A,~U,~S,~P,~\rho,~\beta$ are 
auxiliary fields, so they cannot appear with derivatives in the component action. The rest of the degrees of freedom 
left are exactly those that compose the  half odd superspin supermultiplet and therefore, the action in components 
has to be the Fronsdal action.

To see in details how all this takes place, we substitute the component field expansions from the above expression
for the superfields to the action(\ref{equ32}). The bosonic piece is: 
\be
\eqalign{
S_{Bosons}=\int d^4x&~\[\frac{2(s+1)^2}{s(2s+1)}\]c_3h^{\a(s+1)\ad(s+1)}\Box h_{\a(s+1)
\ad(s+1)}\cr
-&\[\frac{(s+1)^3}{s(2s+1)}\]c_3h^{\a(s+1)\ad(s+1)}\pa_{\a_{s+1}\ad_{s+1}}\pa^{\g\gd}h_{
\g\a(s)\gd\ad(s)}\cr
+&\[\frac{2(s+1)^3}{(2s+1)}\]c_3h^{\a(s+1)\ad(s+1)}\pa_{\a_{s+1}\ad_{s+1}}\pa_{\a_s
\ad_s}h_{\a(s-1)\ad(s-1)}\cr
-&\[\frac{2(s+1)^3}{s}\]c_3h^{\a(s-1)\ad(s-1)}\Box h_{\a(s-1)\ad(s-1)}\cr
-&\[\frac{(s+1)^3(s-1)^2}{s(2s+1)}\]c_3h^{\a(s-1)\ad(s-1)}\pa_{\a_{s-1}\ad_{s-1}}\pa^{
\g\gd}h_{\g\a(s-2)\gd\ad(s-2)}\cr
&+\[\frac{4s}{2s+1}\]c_3 S^{\a(s-1)\ad(s-1)}S_{\a(s-1)\ad(s-1)}\cr
&+\[\frac{4s}{2s+1}\]c_3 P^{\a(s-1)\ad(s-1)} P_{\a(s-1)\ad(s-1)}\cr
&+\[\frac{4(s+1)^3-16s^4}{s(2s+1)}\]c_3A^{\a(s)\ad(s)}{{{A}}}_{\a(s)\ad(s)}\cr
&+4c_3v^{\a(s)\ad(s)}v_{\a(s)\ad(s)}\cr
&-\[\frac{4}{2s+1}\]c_3w^{\a(s)\ad(s)}w_{\a(s)\ad(s)}\cr
&+\[\frac{2(s+1)(s-1)}{s(2s+1)}\]c_3U^{\a(s)\ad(s-2)}U_{\a(s)
\ad(s-2)}+c.c.
}
\ee
The component fields above all correspond to the zero-$\theta$ limit of a corresponding
superfield. The equations of motions for the auxiliary superfields are:
\be
\eqalign{
&{{{A}}}_{\a(s)\ad(s)}=0  ~~~, \cr
&S_{\a(s-1)\ad(s-1)}=0 ~~~,   \cr
&P_{\a(s-1)\ad(s-1)}=0  ~~~, \cr
&v_{\a(s)\ad(s)}=0  ~~~, \cr
&U_{\a(s)\ad(s-2)}=0 ~~~,  \cr
&w_{\a(s)\ad(s)}=0 ~~~.  }
\label{equ43}
\ee
and the final action for the propagating bosonic components is
\be
\eqalign{
S_{Bosons}=\int d^4x&~\[\frac{2(s+1)^2}{s(2s+1)}\]c_3h^{\a(s+1)\ad(s+1)}\Box h_{\a(s+1)
\ad(s+1)}\cr
-&\[\frac{(s+1)^3}{s(2s+1)}\]c_3h^{\a(s+1)\ad(s+1)}\pa_{\a_{s+1}\ad_{s+1}}\pa^{\g\gd}h_{
\g\a(s)\gd\ad(s)}\cr
+&\[\frac{2(s+1)^3}{(2s+1)}\]c_3h^{\a(s+1)\ad(s+1)}\pa_{\a_{s+1}\ad_{s+1}}\pa_{\a_s
\ad_s}h_{\a(s-1)\ad(s-1)}\cr
-&\[\frac{2(s+1)^3}{s}\]c_3h^{\a(s-1)\ad(s-1)}\Box h_{\a(s-1)\ad(s-1)}\cr
-&\[\frac{(s+1)^3(s-1)^2}{s(2s+1)}\]c_3h^{\a(s-1)\ad(s-1)}\pa_{\a_{s-1}\ad_{s-1}}\pa^{
\g\gd}h_{\g\a(s-2)\gd\ad(s-2)}
~~~.}
\label{equ44}
\ee
By setting $c_3=\frac{s(2s+1)}{2(s+1)^2}$ we obtain:
\be
\eqalign{
S_{Bosons}=\int d^4x&~h^{\a(s+1)\ad(s+1)}\Box h_{\a(s+1)
\ad(s+1)}\cr
-&\frac{s+1}{2}h^{\a(s+1)\ad(s+1)}\pa_{\a_{s+1}\ad_{s+1}}\pa^{\g\gd}h_{
\g\a(s)\gd\ad(s)}\cr
+&(s+1)sh^{\a(s+1)\ad(s+1)}\pa_{\a_{s+1}\ad_{s+1}}\pa_{\a_s
\ad_s}h_{\a(s-1)\ad(s-1)}\cr
-&(s+1)(2s+1)h^{\a(s-1)\ad(s-1)}\Box h_{\a(s-1)\ad(s-1)}\cr
-&\frac{(s+1)(s-1)^2}{2}h^{\a(s-1)\ad(s-1)}\pa_{\a_{s-1}\ad_{s-1}}\pa^{
\g\gd}h_{\g\a(s-2)\gd\ad(s-2)}
~~~,}
\ee
which is the Fronsdal action for a propagating spin-($s+1$) bosonic field.  For the
limiting value of $s$ $=$ 1, this is the linearized Einstein-Hilbert action.  

The fermionic piece of the action is:
$$
\eqalign{
S_{Fermions}=\int d^4x &\[\frac{4(s+1)^2}{s(2s+1)}\]ic_3\bar{\psi}^{\a(s)\ad(s+1)}\pa^{
\a_{s+1}}{}_{\ad_{s+1}}\psi_{\a(s+1)\ad(s)}   {~~~~~~~~~~~~}     {~~~~~~~~~~~~~~}   \cr
 } 
$$
\be
\eqalign{
~~~~~~&-\[\frac{2}{s}\]ic_3\bar{\psi}^{\a(s-1)\ad(s)}\pa^{\a_s}{}_{\ad_s}\psi_{\a(s)\ad(s-1)}\cr
&+\[\frac{2(s+1)}{2s+1}\]ic_3\psi^{\a(s+1)\ad(s)}\pa_{\a_{s+1}\ad_{s}}\psi_{\a(s)\ad(s-1)
}+c.c.\cr
&-\[\frac{2(s+1)^2}{s(2s+1)}\]ic_3\psi^{\a(s)\ad(s-1)}\pa_{\a_s\ad_{s-1}}\psi_{\a(s-1)\ad(
s-2)}+c.c.\cr
&-\[\frac{2(s+1)^2}{s(2s+1)}\]ic_3\bar{\psi}^{\a(s-2)\ad(s-1)}\pa^{\a_{s-1}}{}_{\ad_{s-1}}
\psi_{\a(s-1)\ad(s-2)}\cr
&+\[\frac{4(s+1)}{2s+1}\]c_3\beta^{\a(s)\ad(s-1)}\rho_{\a(s)\ad(s-1)}+c.c.\cr
~~~. }
\ee
The equation of motion for the fermionic auxiliary fields are
\be
\rho_{\a(s)\ad(s-1)}=0  ~~~, ~~~\beta_{\a(s)\ad(s-1)}=0
\label{equ47}
\ee
and the action for propagating fermions takes it's final form:
\be
\eqalign{
S_{Fermions}=\int d^4x &\[\frac{2(s+1)^2}{s(2s+1)}\]ic_3\bar{\psi}^{\a(s)\ad(s+1)}\pa^{
\a_{s+1}}{}_{\ad_{s+1}}\psi_{\a(s+1)\ad(s)}\cr
&-\[\frac{2}{s}\]ic_3\bar{\psi}^{\a(s-1)\ad(s)}\pa^{\a_s}{}_{\ad_s}\psi_{\a(s)\ad(s-1)}\cr
&+\[\frac{2(s+1)}{2s+1}\]ic_3\psi^{\a(s+1)\ad(s)}\pa_{\a_{s+1}\ad_{s}}\psi_{\a(s)\ad(s-1)
}+c.c.\cr
&-\[\frac{2(s+1)^2}{s(2s+1)}\]ic_3\psi^{\a(s)\ad(s-1)}\pa_{\a_s\ad_{s-1}}\psi_{\a(s-1)\ad(
s-2)}+c.c.\cr
&-\[\frac{2(s+1)^2}{s(2s+1)}\]ic_3\bar{\psi}^{\a(s-2)\ad(s-1)}\pa^{\a_{s-1}}{}_{\ad_{s-1}}
\psi_{\a(s-1)\ad(s-2)}
~~~. }
\label{equ48}
\ee
Let's set the value for $c_3=\frac{s(2s+1)}{2(s+1)^2}$ as in the bosonic case,
the action becomes 
\be
\eqalign{
S_{Fermions}=\int d^4x~~ &i\bar{\psi}^{\a(s)\ad(s+1)}\pa^{
\a_{s+1}}{}_{\ad_{s+1}}\psi_{\a(s+1)\ad(s)}\cr
&-\[\frac{2s+1}{(s+1)^2}\]i\bar{\psi}^{\a(s-1)\ad(s)}\pa^{\a_s}{}_{\ad_s}\psi_{\a(s)\ad(s-1)}\cr
&+\[\frac{s}{s+1}\]i\psi^{\a(s+1)\ad(s)}\pa_{\a_{s+1}\ad_{s}}\psi_{\a(s)\ad(s-1)
}+c.c.\cr
&-i\psi^{\a(s)\ad(s-1)}\pa_{\a_s\ad_{s-1}}\psi_{\a(s-1)\ad(
s-2)}+c.c.\cr
&-i\bar{\psi}^{\a(s-2)\ad(s-1)}\pa^{\a_{s-1}}{}_{\ad_{s-1}}
\psi_{\a(s-1)\ad(s-2)}
~~~. }
\ee
which is the Fronsdal action for spin-($s+1/2$).
Therefore we conclude only an irreducible supermultiplet propagates on-shell 
and therefore the action(\ref{equ32}) describes a massless half odd superspin
$Y$ $=$ $s$+1/2.
The counting of the off-shell degrees of freedom for this action including all the auxiliary
fields is:
\begin{center}
\begin{tabular} {| c | c | c |}
\hline
Component Field(s) & Bosonic  & Fermionic\\
\hline
$h_{\a(s+1)\ad(s+1)}$~/~$h_{\a(s-1)\ad(s-1)}$ & $s^2  + 2 s  + 3$ & {}\\
\hline
$\psi_{\a(s+1)\ad(s)}$~/~$\psi_{\a(s)\ad(s-1)}$~/~$\psi_{\a(s-1)\ad(s-2)}$  & {} & $4 ( 
s^2  + s + 1 ) $\\
\hline
$A_{\a(s)\ad(s)}$ & $(s+1)^2$ & {}\\
\hline
$\rho_{\a(s)\ad(s-1)}$ & {} & $2 s (s+1) $\\
\hline
$U_{\a(s)\ad(s-2)} $ & $ 2(s+1)(s-1)$ & {}\\
\hline
$v_{\a(s)\ad(s)} $ & $(s+1)^2$ & {}\\
\hline
$w_{\a(s)\ad(s)}$ & $(s+1)^2$ & {}\\
\hline
$S_{\a(s-1)\ad(s-1)}$ & $s^2$ & {}\\
\hline
$P_{\a(s-1)\ad(s-1)}$ & $s^2$ & {}\\
\hline
$\beta_{\a(s)\ad(s-1)}$ & {} & $2 s (s+1) $\\
\hline\hline
$~$ & $8s^2+8s+4$ & $8s^2+8s+4$\\
\hline
\end{tabular}
\end{center}

~~The results in (\ref{equ35}), (\ref{equ36}), and (\ref{equ37}), taken together 
with the component expansions in (\ref{equ39}) and (\ref{equ40}), and the 
component results discussed thereafter are very revealing...when one considers them 
for the special case of the $s$ $=$ 1 theory\footnote{When we take the limit to $s=1$ 
we have to keep in mind, that fields with a negative \newline $~~~~~~$ number of 
undotted or dotted indices must vanish. That is true because these compo-  \newline 
$~~~~~~$ nents do not exist in the $\theta$ expansion of the superfields. Specifically 
the  component \newline $~~~~~~$ $\psi_{\a(s-1)\ad(s-2)}$ which is the
antisymmetric component of the $\theta\bar{\theta}$ term of the $\U$ superfield
\newline $~~~~~~$
doesn't exist in the $s=1$ limit, so it must disappear from the action}.

~~The first component level off-shell description of supergravity was provided in 1977 
in a work by Breitenlohner \cite{nonmin1}.  Two years later and in a subsequent
series of papers \cite{nonmin2}, these results were put into the context of the general
superspace formalism for 4D, $\cal N$ $=$ 1 superfield supergravity.  These old results 
and the special case of the higher spin $s$ $=$ 1 results for B-series discussed above 
match perfectly.  This is especially clear from an examination of the auxiliary fields in
the table immediately above.  In  $s$ $=$ 1 limit, only the $U$ auxiliary boson must
be set to zero and the remaining fields are the well known ones of the non-minimal
off-shell 4D, $\cal N$ $=$ 1 SG multiplet.

Another way to see this, one can initially compare the results of the current paper
for the component expansions given in  (\ref{equ39}) and (\ref{equ40}) to the similar
expansions given in equation (4.9) of the first work in Ref. \cite{nonmin2}.  The 
(\ref{equ35}), (\ref{equ36}), and (\ref{equ37}), can be compared with the linearized
versions of the results found in the remaining papers of Ref. \cite{nonmin2}.  In other 
words, the implication of our present effort reveals that the non-minimal formulation 
of Breitenlohner is the lowest member of a class of arbitrary higher superspin, off-shell 
formulation of massless supermultiplets!


$$
\vCent
 {\setlength{\unitlength}{1mm}
  \begin{picture}(-20,-140)
   \put(-56,-61){\includegraphics[width=4.5in]{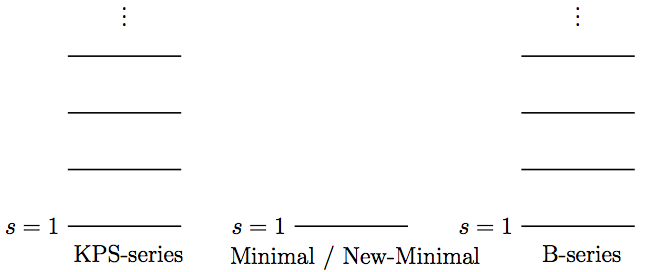}}
  \end{picture}}
  $$
\vskip2.5in
\section{Perspectives On Future Investigations}

$~~~$ In the current work, we have been able to advance the state-of-the-art with
regard to the understanding of 4D, $\cal N $ $=$ 1 superfields and the issue of higher
spin supermultiplets.  The discovery of the B-series of superfield theories, suggests
that many features of off-shell 4D, $\cal N $ $=$ 1 supergravity may well persists
in the cases of higher spin ($s$ $>$ 1).  The gauge transformation law in (\ref{equ03})
for $s$ $=$ 1 is known to define the superspace superconformal group.  It thus seems
reasonable for values of $s$ $\ne$ 1 to use this as a definition of the 4D, $\cal N $ $=$ 1 
superspace superconformal group acting on the entire B-series of theories.  Furthermore, 
there is no obvious reason not to use this to define a 4D, $\cal N $ $=$ 1 superspace 
superconformal group for the KPS-series also.

If it is accepted that the gauge transformation law of $H_{\a(s)\ad(s)}$ defines a 4D, $\cal 
N $ $=$ 1 superspace superconformal group, the second equation in (\ref{equ33}) has an 
obvious interpretation.  In the case of $s$ $=$ 1 limit, the superfield $\U_{\a(s)\ad(s-1)}$ is 
known to constitute a conformal compensator whose functions is to break the 4D, $\cal N $ 
$=$ 1 superspace superconformal group down to the 4D, $\cal N $ $=$ 1 superspace super 
Poincar\' e group.   Once more it is suggestive that this interpretation can be carried over
to the entirety of the B-series and as well to the KPS-series (though the governing equations
for the KPS-series are given by (\ref{equ17})).  For both cases we have verified the existence 
of field strength superfields ${\bm{\cal W}}_{\a(2s+1)}$, $ {\bm{{\cal G}}}_{\a(s)\ad(
s)}$ and ${\bm{{\cal T}}}_{\a(s)\ad(s-1)}$ which occur for both the B-series and the
KPS-series.

In a future work, we will revisit all of these results in the context of a Fock-space formulation.
We conjecture that all the structures we have met in this investigation will likely generalize
to such a formulation.  Should this be the case, then we may have a new avenue to 
ask questions of covariant superstring field theory.  Can there exist a limit of covariant 
superstring field theory which recovers all the structure found in a Fock space extension
of our current work?

${~~~}$ \newline
${~~~~~}$``{\it {Never express yourself more clearly than you are able to think.
}}"${~~~}$ \newline
\newline $~~~~~~~$ -- Niels Bohr
\newline ${~~~}$ \newline

\noindent
{\Large\bf Acknowledgments}

This research was supported in part by the endowment of the John S.~Toll Professorship, the University of Maryland 
Center for String \& Particle Theory, National Science Foundation Grant PHY-0354401.  This work is also supported 
by U.S. Department of Energy (D.O.E.) under cooperative agreement DEFG0205ER41360.  SJG offers additional 
gratitude to the M.\ L.\ K. Visiting Professorship and to the M.\ I.\ T.\ Center for Theoretical Physics for support and
hospitality extended during the undertaking of this work.

\newpage
{\bf \Large {Appendix: Recovering the missing $s=1$ pieces}}

~~~For the purpose of completeness we study separately the case of $s=1$. This limit is special
because in a practical level the index structure of the entire theory gets simplified and \eqref{equ04},
which was a guideline, becomes simpler as well. Furthermore it is known that there exist two dual,
well studied, theories of supergravity, the minimal and the new minimal. We would like to find if and 
how they emerge from our construction. For this purpose we will not bother, with the $s=1$ limits of
the two theories described above. Instead we will search for different routes that could lead to a
consistent theory.

There are a couple of interesting observations that one can make. Using the table above that counts
the off-shell degrees of freedom, in the $s=1$ limit gives the answer twenty. Minimal and new minimal
formulations of supergravity are known to have twelve off-shell degrees of freedom. So it can not be a
limit of the above theories. This means that in our framework there must be a different mechanism that is
capable of generating these theories. We would like to find this.

The second and more important observation is that the superfield $H$ in the s=1 limit includes all the
propagating bosonic and fermionic degrees of freedom need to construct the Fronsdal action. It is very
easy to verify that, just be looking equation \eqref{equ39} and \eqref{equ40}. One can check that in the
$s=1$ limit all the $h$'s and $\psi$'s components are in the taylor expansion of the $H_{\a\ad}$. This
means that the compensator looses one of it's roles, to provide the extra degrees of freedom needed
for the irreducible representation. It's sole purpose purpose now is to guarantee the gauge invariance
of the action. This infers that the set of all compensators must be auxiliary superfields (so their mass
dimensions must be one) and as a result their gauge transformation must be made out of 3 $\D$'s
($\Dd$'s) acting on the only gauge parameter available $L_{\a}$. So we must look for compensators
that transform like

\begin{center}
\begin{tabular}{c c}
\begin{minipage}{1.9in}
\begin{itemize}
\item ~$\Dd^2\D^{\a}L_{\a}$ 
\end{itemize}
\end{minipage}
& 
\begin{minipage}{1.9in}
\begin{itemize}
\item ~$\D^{\a}\Dd^2L_{\a}$
\end{itemize}
\end{minipage}
\\
\begin{minipage}{1.9in}
\begin{itemize}
\item ~$\D^2\Dd_{\ad}L_{\a}$
\end{itemize}
\end{minipage}
& 
\begin{minipage}{1.9in}
\begin{itemize}
\item ~$\Dd_{\ad}\D^2L_{\a}$
\end{itemize}
\end{minipage}
\end{tabular}
\end{center}
The last two possibilities will introduce compensators with exactly the same index structure (and therefore fields content) as the main superfield, that is why we will not allow them. So there are two cases left that correspond at the minimal and new minimal formulations of supergravity.

The starting action is the $s=1$ version of \eqref{equ01} and the change of this action under the gauge transformation
\eqref{equ03} is 

\be
\eqalign{
\delta S=\int d^8z\Bigg\{
&~\[-2c_1+2c_2+2c_3+6c_4\]H^{\a\ad}\Dd_{\ad}
\D^2\Dd^2L_{\a}      \cr  
&+2c_2 H^{\a)\ad}\Dd^2\D^2\Dd_{\ad}L_{\a}\cr
&+\[-2c_3+6c_4\]H^{\a\ad}\D_{\a}\Dd^2\D^{\g}
\Dd_{\ad}L_{\g}\cr
&+\[2c_3-2c_4\]H^{\a\ad}\Dd_{\ad}\D_{\a}\Dd^2\D^{\g}L_{\g}\cr
&+c.c.\Bigg\}    ~~~.
} 
\ee
Setting $c_2=0$ gives

\be
\eqalign{
\delta S=\int d^8z\Bigg\{
&~H^{\a\ad}\Big\{(-2c_3+6c_4)\D_{\a}\Dd_{\ad}-(-2c_1+2c_3+6c_4)\Dd_{\ad}\D_{\a}\Big\}\D^{\g}\Dd^2L_{\g}\cr
&+2(c_3-c_4)H^{\a\ad}\Dd_{\ad}\D_{\a}\Dd^2\D^{\g}L_{\g}\cr
&+c.c.\Bigg\}    ~~~.
} 
\ee

If $-2c_3+6c_4=-2c_1+2c_3+6c_4 \Rightarrow c_1=2c_3$
then
\be
\eqalign{
\delta S=\int d^8z\Bigg\{
&~(-2c_3+6c_4)H^{\a\ad}\[\D_{\a},\Dd_{\ad}\]\Big\{\D^{\g}\Dd^2L_{\g}+\Dd^{\gd}\D^2\bar{L}_{\gd}\Big\}\cr
&+2(c_3-c_4)H^{\a\ad}\Dd_{\ad}\D_{\a}\Big\{\Dd^2\D^{\g}L_{\g}\Big\}+c.c.\Bigg\}    ~~~.
} 
\ee
 
 At this point, we introduce two compensators, a real scalar $U$ with mass dimensions $[U]=1$ and a complex
scalar $\sigma$ with mass dimensions $[\sigma]=1$. Their transformations are defined to be
\be
\eqalign{
&\delta U=\D^{\g}\Dd^2L_{\g}+\Dd^{\gd}\D^2\bar{L}_{\gd}\cr
&\delta \s=\Dd^2\D^{\g}L_{\g}
}
\ee

We also add to the action the following terms so the compensators have dynamics

\be
\eqalign{
S_{c}=\int d^8z\Bigg\{
&~~-(-2c_3+6c_4)H^{\a\ad}\[\D_{\a},\Dd_{\ad}\]U\cr
&~~-2(c_3-c_4)H^{\a\ad}\Dd_{\ad}\D_{\a}\sigma +c.c.\Bigg\}\cr
{}\cr
S_{k.e}=\int d^8z\Bigg\{
&~~bU^2+e\s\bar{\s}+f\s\s+f^*\bar{\s}\bar{\s}\Bigg\}\cr
S_{int.}=\int d^8z\Bigg\{
&~~gU(\s+\bar{\s})\Bigg\}
}
\ee

The full action is
\be
\eqalign{
S=\int d^8z\Bigg\{
~&2c_3 H^{\a\ad}\D^{\g}\Dd^2\D_{\g}H_{\a\ad}\cr
+&c_3 H^{\a\ad}\pa_{\a\ad}\pa^{\g\gd}H_{\g\gd}\cr
+&c_4 H^{\a\ad}[\D_{\a},\Dd_{\ad}][\D^{\g},\Dd^{\gd}]H_{\g\gd}\cr
-&(-2c_3+6c_4)H^{\a\ad}\[\D_{\a},\Dd_{\ad}\]U\cr
-&2(c_3-c_4)H^{\a\ad}\Dd_{\ad}\D_{\a}\sigma +c.c.\cr
+&bU^2+e\s\bar{\s}+f\s\s+f^*\bar{\s}\bar{\s}\cr
+&gU(\s+\bar{\s})\Bigg\}
}
\ee

The invariance of the above action under the gauge transformations gives  the following
Bianchi identity
\be
\Dd^{\ad}\mathcal{G}_{\a\ad}-\Dd^2\D_{\a}\mathcal{E}^1-\D_{\a}\Dd^2\mathcal{E}^2=0
\ee

where $\mathcal{G}_{\a\ad},~\mathcal{E}^1,~\mathcal{E}^2$ are the variations of the full action
with respect the superfields $H_{\a\ad},~U,~\s$

\be
\eqalign{
\mathcal{G}_{\a\ad}=~&4c_3\D^{\g}\Dd^2\D_{\g}H_{\a\ad}+2c_3\pa_{\a\ad}\pa^{\g\gd}H_{\g\gd}\cr
+&2c_4[\D_{\a},\Dd_{\ad}][\D^{\g},\Dd^{\gd}]H_{\g\gd}-(-2c_3+6c_4)\[\D_{\a},\Dd_{\ad}\]U\cr
-&2(c_3-c_4)\Dd_{\ad}\D_{\a}\sigma+(c_3-c_4)\D_{\a}\Dd_{\ad}\bar{\s}\cr
{}\cr
\mathcal{E}^1=~&(2c_3-6c_4)\[\D^{\g},\Dd^{\gd}\]H_{\g\gd}+2bU\cr
+&g\s+g\bar{\s}\cr
{}\cr
\mathcal{E}^2=~&2(c_3-c_4)\D^{\g}\Dd^{\gd}H_{\g\gd}+e\bar{\s}+2f\s\s+gU
}
\ee

The solution of the Bianchi identity is
\begin{center}
\begin{tabular}{c c}
\begin{minipage}{1.9in}
\begin{itemize}
\item $\s$ is chiral
\end{itemize}
\end{minipage}
& 
\begin{minipage}{1.9in}
\begin{itemize}
\item $c_3=c_4$
\end{itemize}
\end{minipage}
\\
\begin{minipage}{1.9in}
\begin{itemize}
\item $e=0$
\end{itemize}
\end{minipage}
& 
\begin{minipage}{1.9in}
\begin{itemize}
\item $b=6c_4$
\end{itemize}
\end{minipage}
\\
\begin{minipage}{1.9in}
\begin{itemize}
\item $g=4c_4$
\end{itemize}
\end{minipage}
\end{tabular}
\end{center}

Hence the action takes the form

\be
\eqalign{
S=\int d^8z\Bigg\{
~~&2c_4 H^{\a\ad}\D^{\g}\Dd^2\D_{\g}H_{\a\ad}\cr
&+c_4 H^{\a\ad}\pa_{\a\ad}\pa^{\g\gd}H_{\g\gd}\cr
&+c_4 H^{\a\ad}[\D_{\a},\Dd_{\ad}][\D^{\g},\Dd^{\gd}]H_{\g\gd}\cr
&-4c_4H^{\a\ad}\[\D_{\a},\Dd_{\ad}\]U\cr
&+6c_4U^2\cr
&+4c_4U(\s+\bar{\s})\Bigg\}
}
\ee
which is invariant under the gauge transformations:
\be
\eqalign{
& \delta H_{\a\ad}=\Dd_{\ad}L_{\a}-\D_{\a}\bar{L}_{\ad}\cr
&\delta U=\D^{\g}\Dd^2L_{\g}+\Dd^{\gd}\D^2\bar{L}_{\gd}\cr
&\delta \s=\Dd^2\D^{\g}L_{\g}
}
\ee
The equation of motion for the superfield $U$ is 
\be
\eqalign{
\mathcal{E}^1=0 \Rightarrow U=\frac{1}{3}\[\D^{\g},\Dd^{\gd}\]H_{\g\gd}-\frac{1}{3}\left(\s+\bar{\s}\right)
}
\ee
and the action becomes:
\be
\eqalign{
S_1=\int d^8z\Bigg\{
~~&2c_4 H^{\a\ad}\D^{\g}\Dd^2\D_{\g}H_{\a\ad}\cr
&+c_4 H^{\a\ad}\pa_{\a\ad}\pa^{\g\gd}H_{\g\gd}\cr
&+\frac{1}{3}c_4 H^{\a\ad}[\D_{\a},\Dd_{\ad}][\D^{\g},\Dd^{\gd}]H_{\g\gd}\cr
&+i\frac{4}{3}c_4H^{\a\ad}\pa_{\a\ad}\left(\bar{\s}-\s\right)\cr
&-\frac{4}{3}c_4\s\bar{\s}\Bigg\}
}
\ee
which is invariant under the transformations:
\be
\eqalign{
& \delta H_{\a\ad}=\Dd_{\ad}L_{\a}-\D_{\a}\bar{L}_{\ad}\cr
&\delta \s=\Dd^2\D^{\g}L_{\g}
}
\ee
The action $S_1$, up to redefinitions,  is the minimal supergravity formulation.

Instead of using the equation of motion for the superfield $U$, we can use the equation of 
motion for the superfield $\s$ then we get:
\be
\mathcal{E}^2=0 \Rightarrow \Dd^2U=0
\ee
therefore $U$ is now a linear compensator and the action is
\be
\eqalign{
S_2=\int d^8z\Bigg\{
~~&2c_4 H^{\a\ad}\D^{\g}\Dd^2\D_{\g}H_{\a\ad}\cr
&+c_4 H^{\a\ad}\pa_{\a\ad}\pa^{\g\gd}H_{\g\gd}\cr
&+c_4 H^{\a\ad}[\D_{\a},\Dd_{\ad}][\D^{\g},\Dd^{\gd}]H_{\g\gd}\cr
&-4c_4H^{\a\ad}\[\D_{\a},\Dd_{\ad}\]U\cr
&+6c_4U^2\Bigg\}
}
\ee
which is invariant under the transformations
\be
\eqalign{
& \delta H_{\a\ad}=\Dd_{\ad}L_{\a}-\D_{\a}\bar{L}_{\ad}\cr
&\delta U=\D^{\g}\Dd^2L_{\g}+\Dd^{\gd}\D^2\bar{L}_{\gd}
}
\ee
This action is the new-minimal supergravity formulation


\end{document}